\documentclass{article}

\usepackage[a4paper, total={6in, 8in}]{geometry}
\usepackage{amsmath}
\usepackage{amsfonts}
\usepackage{mathtools}
\usepackage{comment}
\usepackage{hyperref}

\renewcommand{\author}[1]{{\vspace{5mm}%
   \fontsize{10}{12}
      \raggedright \if@anonymous Author list removed for anonymity \else #1 \fi
	  \vspace{3mm}}}

\newcommand{\affil}[1]{{\fontsize{8}{10}\selectfont
       \raggedright \if@anonymous \phantom{#1} \else #1 \fi}
	   }

\newcommand{\email}[1]{\vspace*{12pt}{\fontsize{8}{10}\selectfont
       \raggedright {\bfseries E-mail:} \if@anonymous \phantom{#1} \else #1 \fi}
	  \vspace{3mm} }

\title{Experimental exclusion of a generalized Károlyházy gravity-induced
decoherence model}
\date{}

\begin{document}

\maketitle

\author{
Nicola Bortolotti$^{1,2,3}$,
Kristian Piscicchia$^{2,3,*}$,
Matthias Laubenstein$^{4}$,
Simone Manti$^{3}$,
Antonino Marcianò$^{5}$,
Federico Nola$^{3,6,7}$,
Catalina Curceanu$^{3,8}$.
}
\\

\affil{$^{1}$Physics Department, “Sapienza” University of Rome, Piazzale Aldo Moro 5, 00185, Rome, Italy}

\affil{$^{2}$Centro Ricerche Enrico Fermi - Museo Storico della Fisica e Centro Studi e Ricerche “Enrico Fermi”, Via Panisperna 89 A, 00184, Rome, Italy}

\affil{$^{3}$Laboratori Nazionali di Frascati, Istituto Nazionale di Fisica Nucleare, Via Enrico Fermi 54, 00044, Frascati, Italy}

\affil{$^{4}$Laboratori Nazionali del Gran Sasso, INFN, Italy}

\affil{$^{5}$Center for Field Theory and Particle Physics \& Department of Physics, Fudan University, 200433, Shanghai, China}

\affil{$^{6}$Dipartimento di Matematica e Fisica, Università degli Studi della Campania “Luigi Vanvitelli”, viale Abramo Lincoln 5, I-81100, Caserta, Italy}

\affil{$^{7}$Istituto Nazionale di Fisica Nucleare, Complesso Universitario di Monte S. Angelo,Via Cintia, I-80126, Napoli, Italy}

\affil{$^{8}$IFIN-HH, Institutul National pentru Fizica si Inginerie Nucleara Horia Hulubei, 30 Reactorului, 077125, M\u agurele, Romania}

%\affil{$^*$Author to whom any correspondence should be addressed.}

\email{kristian.piscicchia@cref.it}

%\keywords{Gravity-related decoherence, Spontaneous wave function collapse, Underground X-ray experiment}

\begin{abstract}
We report new experimental constraints on the generalized version of the
gravity-induced decoherence model originally proposed by Károlyházy.
Using data collected by the VIP Collaboration at the INFN Gran Sasso
National Laboratory with a high-purity germanium detector, we derive an
improved lower bound on the spatial correlation length
$R_K$ characterizing metric fluctuations in the model. We obtain a bound $R_K > 4.64$ m (95\%
C.L.), which exceeds by more than an order of magnitude the previous
experimental limit. When combined with the theoretical upper bound
$R_K <1.98$ m derived from macroscopic localization requirements, our result
excludes the generalized Károlyházy model. The same conclusion applies to an associated non-Markovian formulation of the Continuous Spontaneous Localization (CSL) model.
Our findings significantly tighten
experimental constraints on gravity-related decoherence scenarios and
demonstrate the sensitivity of underground low-background experiments to
foundational modifications of quantum mechanics.%We discuss the implications of this falsification and the prospects for future tests of gravity-related decoherence.
\end{abstract}

%\keywords{Collapse Models; CSL; DP; Spontaneous Radiation; Germanium Detectors.}

%\ccode{PACS numbers:}

%\tableofcontents

\section{Introduction}

This work refines experimental constraints on a generalized version of the gravity-related spontaneous decoherence model \cite{figurato2024testability} originally proposed by Károlyházy \cite{karolyhazy1966,karolyhazy1986possible}. In this framework, a fundamental limit on the precision of space-time length measurements arises from the interplay between quantum mechanics and gravity. This limit manifests as stochastic fluctuations of the metric, which in turn induce decoherence of matter waves. At the same time, metric fluctuations cause spatial diffusion of particles which, if electrically charged, leads to the emission of electromagnetic radiation.

The revised formulation of the Károlyházy model preserves all these key features while relaxing the original assumption that the metric fluctuations must take the form of gravitational waves. As a result, the fluctuations are no longer uniquely constrained. Among the possible realizations, an important class of models is obtained by factorizing the space–time correlation function of the fluctuations into temporal and spatial components: the temporal part is fixed by the space-time uncertainty, whereas the spatial part remains free.

In the non-relativistic limit, the standard Schrödinger equation is modified by a stochastic Newtonian potential. This additional term introduces a fundamental source of irreversibility in the dynamics, leading to decoherence. The corresponding master equation takes a Lindblad form, with the mass density acting as the jump operators. Remarkably, the same master equation also describes the class of mass-proportional spontaneous wave function collapse models \cite{ghirardi1986unified, pearle1989combining, diosi1989models, diosi1987universal, ghirardi1990continuous, ghirardi1990markov} (for comprehensive reviews, see for example \cite{bassi2003dynamical,bassi2013models}), proposed several decades ago as experimentally testable alternatives to standard quantum mechanics that consistently resolve the measurement problem. Although the Károlyházy and spontaneous collapse frameworks generate different evolutions for the wave function—the former linear, the latter non-linear—they yield identical predictions for all possible measurements, rendering them experimentally indistinguishable\footnote{One expects, however, that collapse models in their present formulation are effective descriptions and that a more fundamental underlying theory would lead to some experimentally distinct predictions.}.

In this work, we focus on the case in which the spatial correlation function of the metric fluctuations has a Gaussian profile. This choice allows our experimental analysis to constrain a non-Markovian version of the Continuous Spontaneous Localization (CSL) model, one of the leading collapse model proposals. Notably, the Károlyházy space-time uncertainty fixes both the temporal correlation function and the collapse-rate parameter (usually denoted by $\lambda$), leaving only a single free parameter: the spatial correlation length $R_K$ (corresponding to $R_K=\sqrt{2}r_C$ in standard CSL notation). Considerable effort has been devoted to testing the Markovian form of CSL, in which the noise is totally uncorrelated in time, and the experimentally allowed region of its parameter space has become increasingly restricted. This makes the exploration of more physically realistic, temporally correlated (colored-noise) versions increasingly urgent.

Spontaneous radiation measurements \cite{donadi2021underground,donadi2021novel,majorana} are widely used to test models of spontaneous collapse. Using data from the VIP experiment at the INFN Gran Sasso National Laboratories (LNGS), obtained with a high-purity germanium detector–based setup, we derive an improved lower bound on the characteristic correlation length $R_K$. Our bound surpasses the limit reported in Ref. \cite{figurato2024testability} by more than an order of magnitude.

Ref. \cite{figurato2024testability} also presents a gedanken experiment that provides an upper bound on
$R_K$ by requiring that a macroscopic object—specifically, a single-layer
graphene disk of diameter 10 $\mu \text{m}$ (approximately the smallest size visible to the human eye)—be spatially localized within a time interval of $t = 0.01$ s (the smallest time resolution perceivable by the human eye). Combining our lower bound with this theoretical upper bound shows that the Gaussian version of the generalized Károlyházy model cannot simultaneously satisfy both constraints and is therefore excluded. The same conclusion applies to the associated non-Markovian CSL model.

Summarizing, in this work, we close the remaining window experimentally, based on low-background data from the VIP germanium detector
operated underground at LNGS, and a Bayesian spectral analysis that
incorporates detailed material-dependent emission rates. Our new bound 
excludes the generalized Károlyházy model \cite{figurato2024testability} and the associated non-Markovian CSL model.

Section \ref{brief} reviews the original and generalized forms of the Károlyházy and presents the spontaneous radiation rate for a generic material. Section \ref{exp} describes the experimental apparatus, and Section \ref{dataana} reports the data analysis and results.

\section{Brief review of the Károlyházy model and its generalization}\label{brief}

The core idea of the Károlyházy model is that quantum mechanics imposes a fundamental limit on the precision with which length measurements can be made. This limit is interpreted as an inherent \emph{haziness} in the space-time metric. As a result, matter experiences fluctuations in space-time, and its wave function decoheres when quantized.

Károlyházy's model begins with the observation that the precision with which the time-like length $cT$ can be measured, in flat space-time, by a quantum probe obeying the Heisenberg uncertainty principle (HUP) is finite. By imposing the HUP and the condition that the system's mass is sufficiently small to ensure that its Schwarzschild radius is smaller than the uncertainty in position, the minimum resolution is derived as follows
\begin{equation}\label{uncertainty1}
(\Delta s)^3 = l_p^2 \, s,
\end{equation}
where $l_p$ is the Planck length. This fundamental limitation is interpreted as arising from the stochastic fluctuations in the flat Minkowski metric, given by $(g_{\mu\nu})_\beta(\textbf{x},t) = \eta_{\mu\nu}(\textbf{x},t) + (h_{\mu\nu})_\beta(\textbf{x},t)$, where the different realizations of the metric are indexed by $\beta$. Consequently, the proper length becomes a stochastic variable, with a mean value of $cT$ and a dispersion determined by Eq. \eqref{uncertainty1}. Averaging over the stochastic metric, the mean and dispersion characterize the noise. Under the assumptions of non-relativistic dynamics and weak gravitational fields, only the $g_{00}$ component of the metric is relevant
\begin{equation}
(g_{00})_\beta(\textbf{x},t) = 1 + \gamma_\beta(\textbf{x},t) , 
\end{equation}
where $\gamma_\beta$ represents the noise. The correlation function for this noise is derived assuming that $\gamma_\beta(\textbf{x},t)$ satisfies the wave equation $\Box \gamma_\beta=0$, and its Fourier expansion is performed. Károlyházy assumes that the different modes are independent, and that the correlation depends only on a function $F(k)$ of the modulus of the wave vector, i.e.
\begin{equation}\label{correlation1}
\mathbb{E}\left[ c_\beta(\mathbf{k}) c_\beta^*(\mathbf{k}') \right] = \delta_{\mathbf{k}\mathbf{k}'} \, F(k).
\end{equation}
From the position uncertainty \eqref{uncertainty1}, the correlation function is obtained
\begin{equation}
F(k) = \frac{8\pi^2}{3\Gamma(1/3)} \, l_p^{4/3}\, k^{-5/3},
\end{equation}
with a high-energy cutoff assumed at $k < 2 \pi/\lambda_c$, where $\lambda_c = 10^{-15}$ m.

A pioneering analysis of the \emph{spontaneous radiation} induced by the Károlyházy decoherence mechanism was performed in Ref. \cite{diosi1993calculation}. The space-time haziness corresponds, in the non-relativistic and Newtonian limit, to interaction with a stochastic potential $V_\beta(\textbf{x},t)$ that perturbs the standard Schrödinger evolution
\begin{align}
V_\beta(x,t) = \frac{1}{2}mc^2\gamma_\beta(x,t).
\label{eq_Vbeta}
\end{align}
The associated stochastic acceleration is given by
\begin{align}
\mathbf{a}(\mathbf{x}, t) = -\frac{1}{2} c^2 \nabla \gamma_\beta(\mathbf{x}, t) ,
\label{eq_accel}
\end{align}
which leads to the emission of electromagnetic radiation by charged particles—an effect that is absent in standard quantum mechanics. Ref. \cite{diosi1993calculation} shows that the Károlyházy model predicts an unrealistically high rate of spontaneous radiation, which is inconsistent with experimental observations.

A recent attempt to reconcile the Károlyházy mechanism with experimental measurements of spontaneous radiation rates is provided in \cite{figurato2024testability}. This work highlights that contemporary models of spontaneous wave function collapse, such as the DP and CSL models \cite{diosi1987universal,diosi1989models,ghirardi1986unified,ghirardi1990continuous}, predict a much lower spontaneous radiation rate. This difference is attributed to the distinct structure of the Károlyházy noise correlation function, which is strongly non-Markovian and does not factorize spatial and temporal components, in contrast to the DP and CSL models.

In the same work, a generalization of the Károlyházy model is proposed, relaxing the assumption that $\gamma_\beta$ satisfies the wave equation. Instead, it is assumed that the correlation function for metric fluctuations, $\mathcal{C}^{(\gamma)}(\mathbf{x},t;\mathbf{x}',t') = \mathbb{E}[\gamma_\beta(\mathbf{x},t)\gamma_\beta(\mathbf{x}',t')]$, is translationally invariant in both space and time, as well as invariant under time inversion (i.e., of the form $\mathcal{C}^{(\gamma)}(\mathbf{x}-\mathbf{x'}, |t-t'|)$), with the spatial and temporal parts factorizing 
\begin{equation}
    \mathcal{C}^{(\gamma)}(\mathbf{x}-\mathbf{x'}, |t-t'|) = \mathcal{D}^{(\gamma)}(\mathbf{x}-\mathbf{x'})\mathcal{G}(|t-t'|) .
\end{equation} 
The temporal component is constrained by Eq. \eqref{uncertainty1}, which yields in Fourier space the expression 
\begin{equation}\label{Fourier time correlation}
	\mathcal{G}(\omega) = \frac{4}{\sqrt{3}} \frac{\Gamma\left(2/3\right) t_p^{4/3}  \omega^{1/3}}{\mathcal{D}^{(\gamma)}(0)},
\end{equation}
where $t_p = l_p/c$ is the Planck time, while the spatial component is assumed to be Gaussian with a correlation length $R_K$ that becomes a free parameter in the model. 

In terms of the Newtonian field $\phi_\beta = c^2\gamma_\beta/2$, the spatial part of the correlation function is 
\begin{equation}\label{newtonian correlation}
	\mathcal{D}^{(\phi)}(r) = \frac{c^4t_p}{4}e^{-r^2/R_K^2} .	
\end{equation}
Notably, this model is experimentally equivalent to CSL with non-Markovian noise, where the collapse rate parameter is fixed by the Planck time as
\begin{equation}\label{CSL collapse rate}
	\lambda = \frac{t_p m_0^2c^4}{4\hbar^2} = 27.5 \text{ kHz} ,	
\end{equation}
with $m_0$ the nucleon mass, leaving $R_K$ as the only free parameter. Indeed, the master equation governing CSL dynamics can be derived \cite{bortolotti2025fundamental} from a stochastic Schr\"odinger equation, where the noise field is identified with a stochastic Newtonian potential characterized by zero mean and spatial correlations
\begin{equation}\label{CSL correlation}
    \mathcal{D}_\text{CSL}(r) = \frac{\hbar^2 \lambda}{m_0^2} e^{-r^2/2r_C^2} ,
\end{equation}
where $r_C$ sets the localization resolution of the collapse process. Comparing Eq. \eqref{CSL correlation} with Eq. \eqref{newtonian correlation} then yields expression \eqref{CSL collapse rate} for $\lambda$.

\begin{figure}[h!]
\centering
\includegraphics[width=12cm]{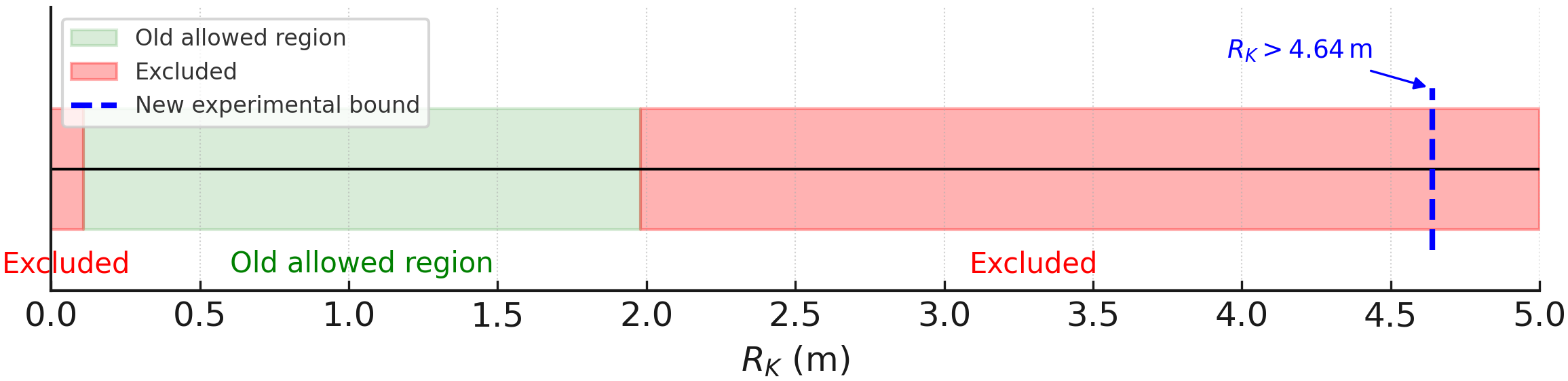}
\caption{The previously allowed region for the correlation length parameter of the generalized k\'arolyh\'azy model, according to the analysis \cite{figurato2024testability} is shown in green. The updated bound $R_K > 4.64$ m obtained in this work is also shown. This new lower value belongs to the forbidden region, thus falsifying the generalized k\'arolyh\'azy mechanism.}
\label{rk}
\end{figure}

In \cite{figurato2024testability}, the spontaneous emission rate for a free particle in the context of the generalized Károlyházy model is derived, yielding
\begin{equation}\label{ratefree}
\frac{d\Gamma}{dE} = \frac{\Gamma(2/3) \alpha c^2 t_p^{4/3}}{\sqrt{3} \pi \hbar^{1/3} R_K^2 E^{2/3}},
\end{equation}
with $\alpha$ the fine structure constant. This result is compared with a recent experimental analysis \cite{majorana}, which provides a lower bound on the correlation length $R_K \ge 0.11$ m. Additionally, as noted in the introduction, a theoretical argument is used to provide an upper bound of $R_K \le 1.98$ m. These results, presented in Ref. \cite{figurato2024testability}, suggest that the Károlyházy model can be compatible with spontaneous radiation measurements, but only within a narrow range of values for $R_K$, as shown in Figure \ref{rk}.

In order to improve sensitivity in the interpretation of the data, we generalized the rate in Eq. \eqref{ratefree} by introducing the effect induced by the emission of the protons and electrons of the atoms of a system. The expression for the rate is equal to that predicted by CSL with noise characterized by the time component \eqref{Fourier time correlation}. For white-noise CSL, the rate emitted by a generic material has been derived in \cite{donadi2021novel} (see Eq. $(5)$); the colored-noise rate is then obtained \cite{piscicchia2024x} by multiplying this expression by the function $\mathcal{G}(\omega)$ given in Eq. \eqref{Fourier time correlation}. Therefore, using expression \eqref{CSL collapse rate} for the collapse rate parameter, we find
\begin{equation}\label{rateatom}
\frac{d\Gamma}{dE} = N_{at} \, (N_p^2 + N_e) \frac{\Gamma(2/3) \alpha c^2 t_p^{4/3}}{\sqrt{3} \pi \hbar^{1/3} R_K^2 E^{2/3}},
\end{equation}
where $N_{at}$ represents the number of atoms,  $N_p$ and $N_e$ the number of protons and electrons, respectively.

\section{Experimental apparatus}\label{exp}

The experimental setup, operated at LNGS, was designed to detect a faint signal of spontaneous radiation emission from the germanium crystal in the detector, as well as from the surrounding materials in the experimental apparatus. The low-background environment at LNGS is ideal for high-sensitivity measurements of extremely low-rate physical processes, as it is shielded by a rock overburden that reduces cosmic muon flux by about six orders of magnitude \cite{donadi2021underground}.

\begin{figure}[h]
    \centering
    \includegraphics[width=0.5\linewidth]{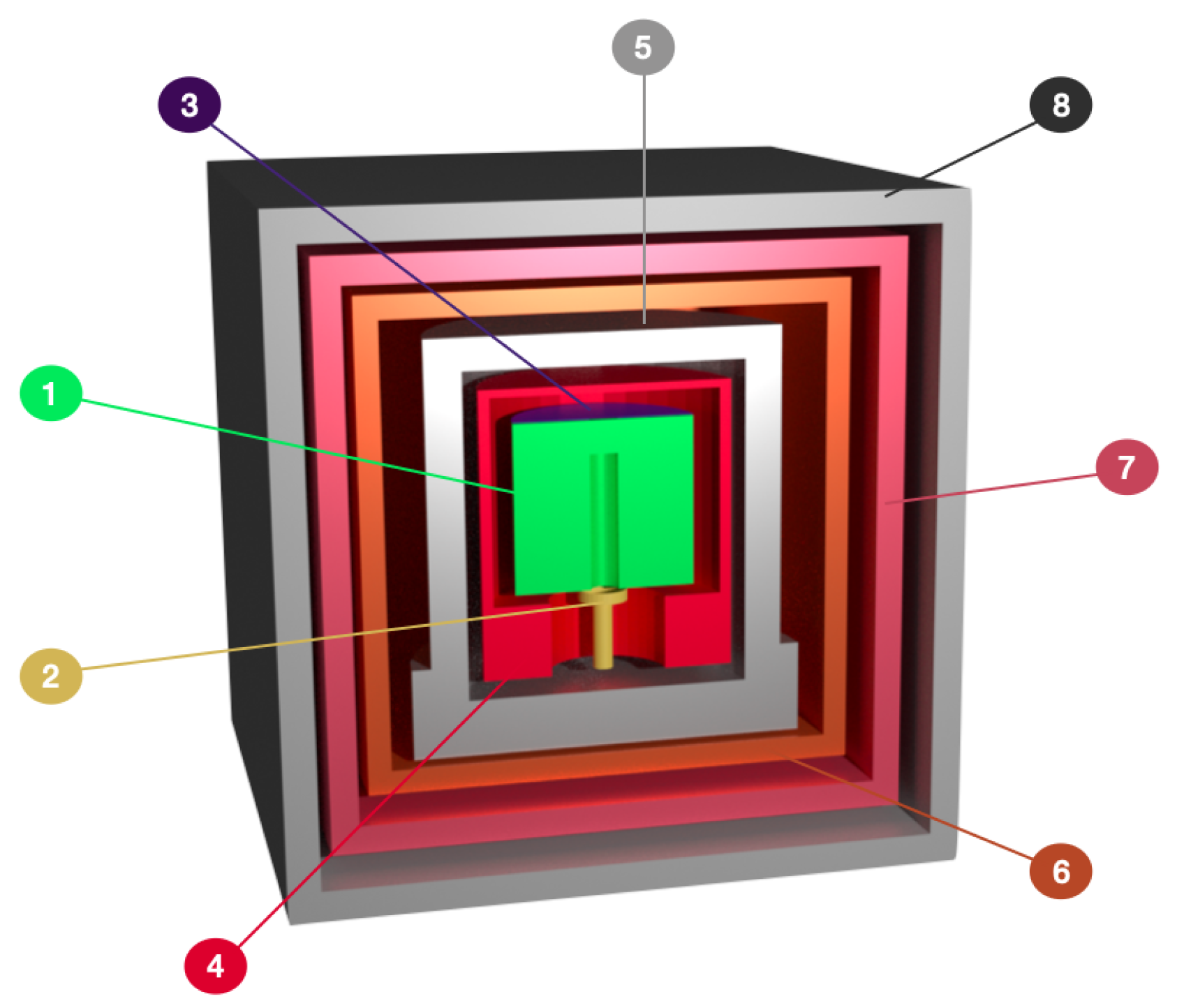}
    \caption{Schematic render of the germanium detector. From inside out, (1) germanium crystal, (2) electric contact, (3) plastic insulator, (4,5,6) Copper cup, end-cup and block, (7) inner Copper shield and (8) Lead shield.}
    \label{fig:viplead}
\end{figure}

The setup, schematically shown in Fig.~\ref{fig:viplead}, consists of a coaxial p-type high-purity germanium detector with a diameter of 8.0 cm and a length of 8.0 cm. It is surrounded by a passive shielding structure, where the outer layer is made of pure lead and the inner layer is composed of electrolytic copper. The active germanium volume of the detector is 375 cm$^3$.

The sample chamber has a volume of approximately 15 L. The shield, along with the cryostat, is enclosed in an air-tight steel housing with a thickness of 1 mm. This housing is continuously flushed with boil-off nitrogen from a liquid nitrogen storage tank to minimize contact with external air, thereby suppressing background from radon $\alpha$ decay.

\begin{figure}[t]
    \centering
    \includegraphics[width=0.75\linewidth]{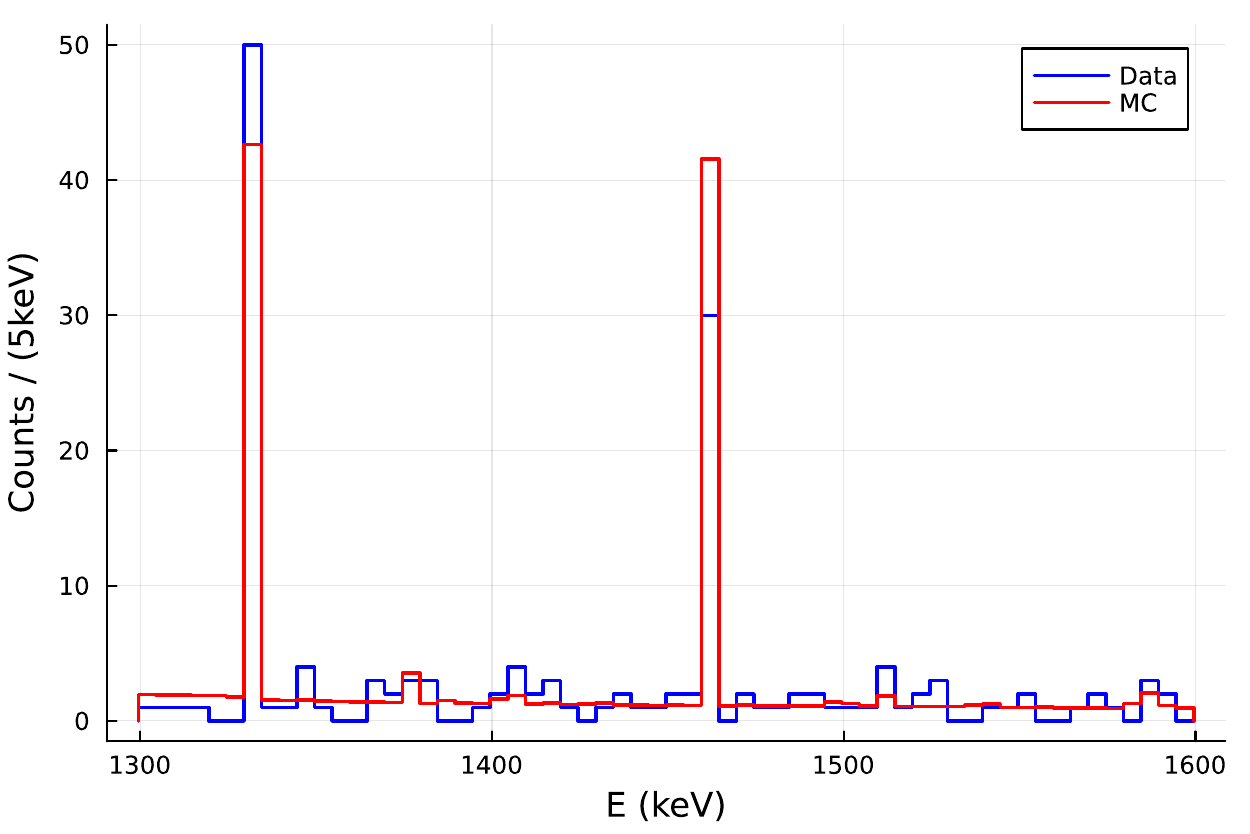}
    \caption{The measured spectrum is represented as a blue distribution. The MC distribution is also shown in red.}
    \label{data}
\end{figure}

The measured emission spectrum, corresponding to a data-taking period of approximately 62 days (from August 2014 to August 2015), and an exposure period of 124 kg $\cdot$ day, is shown in Fig.~\ref{data}. The energy range for the analysis, from $(1.3 - 1.6)$ MeV, was optimized based on a Monte Carlo (MC) study aimed at modeling the background contribution from known emission processes. To fully characterize the experimental setup, calculate detection efficiencies, and simulate the background, the entire detector was incorporated into a validated Monte Carlo (MC) code (Ref.~\cite{boswell2011}), based on the GEANT4 software library (Ref. \cite{agostinelli2003geant4}).

The MC simulation of the background takes as input the measured activities of each radionuclide detected in every component of the setup. The simulation accounts for the emission probabilities and decay schemes, and the photon propagation and interactions within the detector materials, which contribute to the continuum part of the background spectrum, are realistically modeled. Detection efficiencies are also considered in the simulation. The background spectrum is shown in red in Fig. \ref{data}. The measured and simulated spectra are in good agreement in the energy range from $(1.3 - 1.6)$ MeV, where the simulation accounts for more than 97\% of the measured spectrum.
Larger deviations between the measured and simulated spectra are observed in the lower energy region. These discrepancies are primarily due to the inability to perfectly model the residual cosmic rays and the bremsstrahlung produced by ${}^{210}$Pb and its daughters in the massive lead shield. Above 1600 keV, the spectrum becomes sparsely populated and does not provide additional useful information for the bin-by-bin Bayesian analysis described in Section \ref{dataana}.

All assumptions underlying the calculation that leads to the theoretical spontaneous emission rate in Eq.~(\ref{rateatom}) are satisfied within $\Delta E$, provided that the dominant contribution from protons is considered, as electrons become relativistic in this energy range.

\section{Data analysis and results}\label{dataana}

The analysis is aimed at obtaining the probability density function ($pdf$) for the yield of spontaneous radiation ($Y=1/R^2_K$), which potentially contribute to the measured spectrum. As described below, no significant signal is observed. Given that $Y$ is a monotonic function of $R_K$, the $pdf$s of the two stochastic variables are related by the probability invariance under change of variables, and an upper limit on $Y$ translates into a lower limit on $R_K$.

The posterior $pdf$ is provided by the Bayes theorem
\begin{equation}\nonumber
    P(Y| data) = \frac{P(data | Y) \, P_0 (Y)) }{\int P(data | Y) \, P_0 (Y)  dY} \, ,
    \label{posterior}
    \end{equation}
and the likelihood is taken as a product of Poissonian distributions
\begin{equation}
 P(data | Y) = \prod_{i=1}^{N} \frac{\lambda_i (Y)^{n_i} \, e^{-\lambda_i (Y)}}{n_i !} \, ,
\end{equation}
where $n_i$ are the measured bin contents and the expectation values are parametrized as
\begin{equation}
    \lambda_i(Y)=b_i^{MC} + \int_{\Delta E_i} f_S (E,Y) \, dE.
\end{equation}
The background counts in the $i$-th bin, which are derived by the MC simulation, are represented by $b_i^{MC}$. $\Delta E_i$ is the energy range corresponding to the $i$-th bin. $f_S (E,Y)$ is the expected shape of the signal distribution.

\subsection{Expected signal shape and prior on $Y$}

The shape of the signal distribution is obtained by weighting the theoretical expected emission rate, which we rewrite as
\begin{equation}\label{rateprot}
\begin{split}
		&\frac{d\Gamma}{dE} = Y\frac{N_{at} N_p^2 \beta_K}{E^{2/3}} ,\\
		&\beta_K \coloneqq \frac{\Gamma(2/3) \alpha c^2 t_p^{4/3}}{\sqrt{3} \pi \hbar^{1/3}} ,
\end{split}
\end{equation}
with the detection efficiency functions, for the various materials of the experimental setup (labelled by $i$), summing over $i$ and multiplying by the total acquisition time ($T_\text{exp}$)
\begin{equation}\label{fs}
	f_S(E, Y) = \frac{Y T_\text{exp} \beta_{K}}{E^{2/3}} \sum_i\alpha_i \epsilon_i(E) .
\end{equation}
The efficiency functions are obtained by means of polynomial fits to the corresponding MC simulated efficiency distributions
\begin{equation}
	\epsilon_i(E) = \sum_{j=0}^{d_i} \xi_{ij} E^j,
\end{equation}
where $\xi_{ij}$ stands for the matrix of the fit parameters and $d_i$ for the degree of the polynomial expansion. The factors $\alpha_i$ are specific of the material under analysis and are given by
\begin{equation}\label{alpha}
    \alpha_i = m_i \, n_i \, N_{pi}^2.  
\end{equation}
In Eq. \eqref{alpha} $m_i$ is the mass of the setup component, $n_i$ is the number of atoms per unit mass, $N_{pi}$ is the number of protons in each atom type of the material. The coefficients $\xi_{ij}$ which are obtained from the best polynomial fits to the corresponding efficiency distributions are shown in Table \ref{tab:fit_parameters}, with the corresponding statistical errors.

Considered the a priori ignorance on the sensitivity of the current measurement, which could deliver a stronger lower bound than the previous theoretical analysis \cite{figurato2024testability}, as well as a weaker lower bound, we opt for a uniform prior over the physical parameter
\begin{equation}
    [R_K^{\mathrm{min}},R_K^{\mathrm{max}}] = [10^{-10}, 10^{9}] \, \mathrm{m}.
\end{equation}

\begin{table}[t]
    \centering
    \caption{The table summarises the parameters obtained from the best fit to the 
efficiency spectra, for each component of the setup which gives a significant 
contribution. The statistical errors on the parameters are shown as well.}
    \resizebox{\columnwidth}{!}{%
    \begin{tabular}{c|ccccc}
        \hline
        $\boldsymbol{i=}$ & \textbf{Ge crystal} & \textbf{Inner Cu}& \textbf{Cu block + plate} & \textbf{Cu shield chamber} & \textbf{Pb shield}\\
        \hline
        $\xi_{i0}$ &  (4.82 $\pm$ 0.03) $\cdot 10^{-1}$  &  (3.77 $\pm$ 0.04) $\cdot 10^{-2}$ & (2.6 $\pm$ 0.1) $\cdot 10^{-3}$  & (-1.01 $\pm$ 0.07) $\cdot 10^{-5}$ & (-5.76 $\pm$ 0.03) $\cdot 10^{-4}$  \\ 
        $\xi_{i1}$ &  (-4.42 $\pm$ 0.03) $\cdot 10^{-4}$ & (-2.48 $\pm$ 0.03) $\cdot 10^{-5}$  & (2.9 $\pm$ 1.4) $\cdot 10^{-7}$ &  (7.8 $\pm$ 0.1) $\cdot 10^{-8}$ & (3.812 $\pm$ 0.003) $\cdot 10^{-6}$ \\ 
        $\xi_{i2}$ &  (2.10 $\pm$ 0.01) $\cdot 10^{-7}$ & (1.03 $\pm$ 0.01) $\cdot 10^{-8}$  &  (-3.1 $\pm$ 0.5) $\cdot 10^{-10}$ &  (-2.07 $\pm$ 0.06) $\cdot 10^{-11}$ &  (-2.728 $\pm$ 0.001) $\cdot 10^{-9}$ \\
        $\xi_{i3}$ &  (-4.87 $\pm$ 0.03) $\cdot 10^{-11}$ &  (-2.24 $\pm$ 0.04) $\cdot 10^{-12}$ & (5.7 $\pm$ 1.6) $\cdot 10^{-14}$ & (1.61 $\pm$ 0.09) $\cdot 10^{-15}$ &  (9.036 $\pm$ 0.004) $\cdot 10^{-13}$ \\
        $\xi_{i4}$ &  (4.32 $\pm$ 0.07) $\cdot 10^{-15}$  & (1.93 $\pm$ 0.08) $\cdot 10^{-16}$  & (-3.1 $\pm$ 3.3) $\cdot 10^{-18}$  &  - &  (-1.477 $\pm$ 0.001) $\cdot 10^{-16}$ \\
        $\xi_{i5}$ &  -  & -  & -  & -  & (9.60 $\pm$ 0.02) $\cdot 10^{-21}$ \\
        \hline
    \end{tabular}
    }
    \label{tab:fit_parameters}
\end{table}

\subsection{Posterior distributions and novel bound on the correlation length of the generalized Károlyházy model}

\begin{figure} [t]
\centering
\includegraphics[width=10cm]{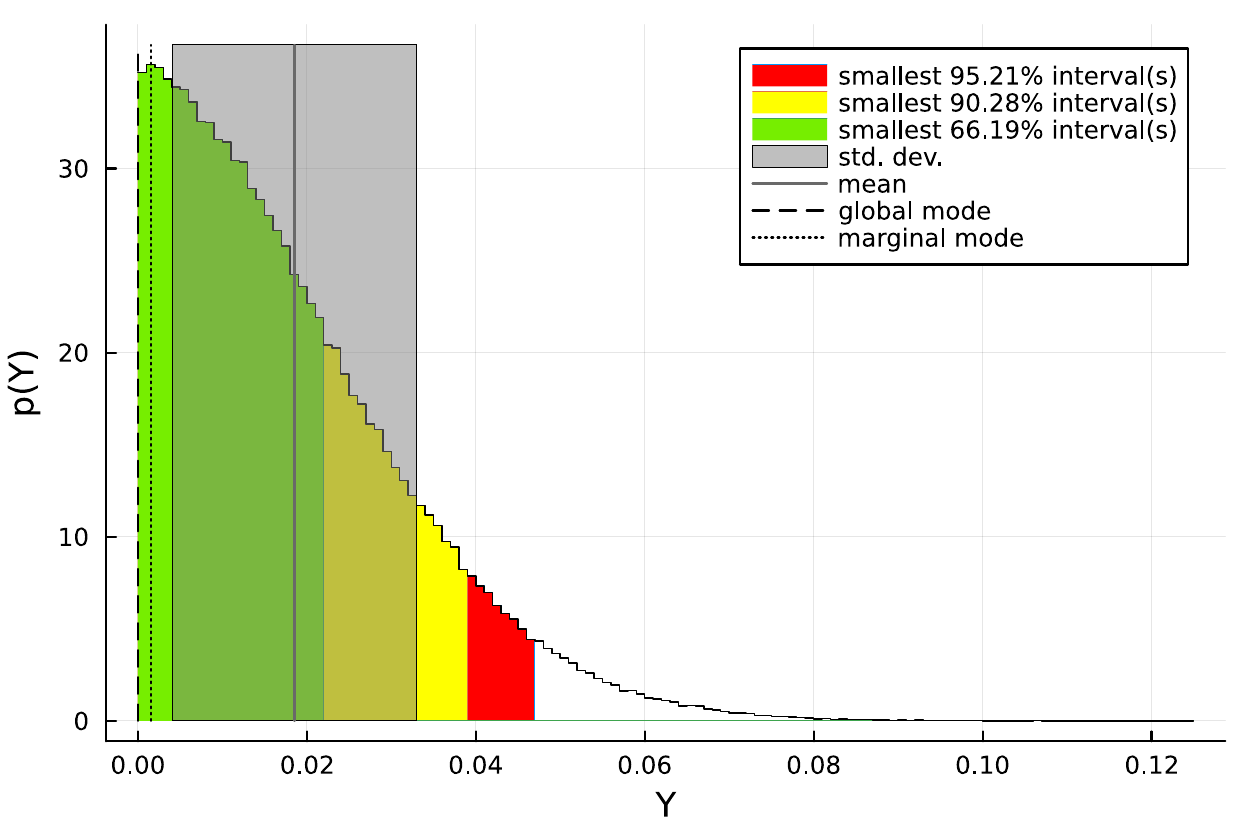}
%\vspace{-0.9cm}
\caption{Posterior probability density functions for $Y = 1/R_K^2$. A dotted line indicates the local mode, the dashed line shows the global mode. The green, yellow and red areas represent the ranges corresponding to probabilities of 0.66, 0.90 and 0.95 respectively.}
\label{fig: posterior}
\end{figure}

The posterior \textit{pdf} in Eq. \eqref{posterior} is shown in Figure \ref{fig: posterior}. Upper limits $\mathcal{L}$ on the parameter $Y$ are then obtained from the cumulative distribution, $\tilde{P}(\mathcal{L})$, by solving the following integral equation for a given probability $\Pi$
\begin{equation}\label{cumulative}
        \tilde{P}(\mathcal{L}) = \int_0^{\mathcal{L}} P(Y | data ) \, dY = \Pi.
\end{equation}
Numerical integrations are performed using BAT, a high-performance toolbox for Bayesian inference \cite{bat1}. The posterior distribution is calculated by means of a Markov Chain Monte Carlo technique, with 4 chains and $10^6$ samples. The following limit is obtained on $R_K$, corresponding to a probability $\Pi = 0.95$
\begin{equation}\label{limit}
   R_K > 4.64  \, \mathrm{m},
\end{equation}
this new bound is also shown in the exclusion region of Figure \ref{rk}. Considered the theoretical lower bound $R_K < 1.98$ m, we can exclude the generalized Károlyházy model developed in \cite{figurato2024testability}.

\section{Conclusion}

In this work, we have performed a high-sensitivity search for
spontaneous radiation predicted by the generalized Károlyházy
gravity-induced decoherence model. We focused on a noise component of the metric tensor with Gaussian spatial correlations and analyzed data collected with the VIP germanium-detector setup at LNGS. By incorporating a refined evaluation
of the emission rate—including explicit contributions from protons and
electrons within the materials of the apparatus—and by performing a
detailed Bayesian analysis of the measured spectrum, we derive a new
lower bound on the spatial correlation length of the model, $R_K > 4.64$ m
(95\% C.L.).

This limit improves previous constraints by more than an order of
magnitude. When combined with the theoretical requirement that the
collapse dynamics must efficiently localize macroscopic systems, leading
to an upper bound $R_K < 1.98$ m, our result rules out the entire remaining
parameter space of the generalized model. We therefore conclude that the
Károlyházy gravity-induced decoherence mechanism, even in its extended Gaussian
form, is experimentally falsified.

Although the generalized Károlyházy framework does not introduce a wave function reduction mechanism, its experimental predictions coincide with those of spontaneous wave function collapse models. In particular, the model analyzed here is experimentally equivalent to a non-Markovian version of CSL. The application of the Károlyházy space-time uncertainty fixes both the temporal correlations—occurring on time scales comparable to the Planck time—and the decay-rate parameter. Our experimental analysis therefore constrains this model as well, ruling it out.

More generally, our results demonstrate how precision measurements developed for rare-event searches—such as VIP—can provide decisive tests of foundational modifications of quantum theory, including models motivated by quantum gravity. Continued advances in detector technology and background suppression are expected to enable even more stringent tests of broader classes of non-Markovian collapse models, gravity-induced decoherence, and alternative quantum-gravity scenarios in the near future.

\section*{Acknowledgments}

This publication was made possible through the support of Grant 62099 from the John Templeton Foundation. The opinions expressed in this publication are those of the authors and do not necessarily reflect the views of the John Templeton Foundation.
We acknowledge support from the Foundational Questions Institute and Fetzer Franklin Fund, a donor advised fund of Silicon Valley Community Foundation (Grants No. FQXi-RFP-CPW-2008 and FQXi-MGA-2102), and from the H2020 FET TEQ (Grant No. 766900).
We thank: the INFN Institute, for
supporting the research presented in this article and, in particular, the Gran Sasso underground laboratory of INFN, INFN-LNGS, and its Director, Ezio Previtali, the LNGS staff, and the Low Radioactivity laboratory for the experimental activities dedicated to the search for spontaneous radiation.
We thank the Austrian Science Foundation (FWF) which supports the VIP2 project with the grants P25529-N20, project P 30635-N36 and W1252-N27 (doctoral college particles and interactions).
K.P. acknowledges support from the Centro Ricerche Enrico Fermi - Museo Storico della Fisica e Centro Studi e Ricerche ``Enrico Fermi'' (Open Problems in Quantum Mechanics project).

\bibliographystyle{unsrt}
\bibliography{reference}

\end{document}